\pdfoutput=1

\documentclass[a4paper]{article}
\usepackage{a4wide}
\usepackage{graphicx}
\usepackage[T1]{fontenc}
\usepackage[utf8]{inputenc}
\usepackage[UKenglish]{babel}
\usepackage{placeins}
\usepackage[shortlabels]{enumitem}
\usepackage[binary-units=true]{siunitx}
\usepackage{color}
\usepackage[dvipsnames]{xcolor}
\usepackage{authblk}
\usepackage{amsmath}
\usepackage{amssymb}
\usepackage{amsthm}
\usepackage{mathtools}
\usepackage{bbold}
\usepackage{dsfont}
\usepackage{braket}
\usepackage[ruled,lined]{algorithm2e}
\usepackage{nicefrac}
\usepackage[backend=bibtex8,style=phys,sorting=none,citestyle=numeric-comp,eprint=true,isbn=true]{biblatex}
\usepackage[pdftex,
pdftitle={Algorithms},
pdfauthor={Johann Ostmeyer},
bookmarks,
linkcolor=Blue,
urlcolor=blue,
colorlinks]{hyperref}
\usepackage{cleveref}

\graphicspath{{figures/},{data/}}

\crefformat{footnote}{#2\footnotemark[#1]#3}

\DeclareMathOperator{\imag}{Im}

\DeclareMathOperator{\tr}{Tr}

\DeclareMathOperator{\diag}{diag}

\DeclareMathOperator{\ord}{\mathcal{O}}
\newcommand{\pdagger}{{\phantom{\dagger}}}

\newcommand{\del}[2]{\ensuremath{\frac{\partial #1}{\partial#2}}}

\newcommand{\eto}[1]{\ensuremath{\mathrm{e}^{#1}}}
\newcommand{\trans}{\ensuremath{\mathsf{T}}}
\newcommand{\md}{\ensuremath{\mathrm{d}}}

\newcommand{\ordnung}[1]{\ensuremath{\ord\left(#1\right)}}
\newcommand{\erwartung}[1]{\ensuremath{\left\langle#1\right\rangle}}
\newcommand{\br}[1]{\ensuremath{\left(#1\right)}}
\newcommand{\brr}[1]{\ensuremath{\left[#1\right]}}

\DeclarePairedDelimiter\abs{|}{|}

\colorlet{myorange}{orange}
\colorlet{mypurple}{purple}
\definecolor{myblue}{HTML}{1E88E5}
\definecolor{mygreen}{HTML}{004D40}

\theoremstyle{definition}

\theoremstyle{remark}

\newcommand{\bonn}{
	\textit{\footnotesize Helmholtz-Institut f\"{u}r Strahlen- und
		Kernphysik,
		University of Bonn, 53115 Bonn, Germany}
}

\makeatletter%
\makeatletter%
\begin{document}
	
	\title{Stable and Efficient Algorithms\\ for the Fermion Determinant}
	
	\author{Johann Ostmeyer\footnote{\href{mailto:ostmeyer@hiskp.uni-bonn.de}{ostmeyer@hiskp.uni-bonn.de}}}
	\affil{\bonn}
	\date{\vspace*{-2\baselineskip}}
	\maketitle
	
	\begin{abstract}
		Some algorithms for the numerically exact treatment of fermion determinants are summarised.
		This is not supposed to be a review, rather a concise handbook.
		The audience is expected to have a basic understanding of how to put fermions on a computer.
		We primarily discuss different ways to work with the fermion matrix in the ``sausage'' (Green's function) formulation for quantum Monte Carlo (QMC).
		We emphasise the need for varied approaches in different space-time volume regimes.
		In particular, for small spatial volumes we describe a numerically stable method based on dense matrix operations.
		It is designed specifically to deal with very low temperature regimes.
		On the other hand, for (relatively) large volumes we describe a highly efficient and scalable sparse matrix approach.
		
		These notes might be expanded in the future.
		As of now, they are not intended to be published in a journal.
		None of the algorithms are new, they are mostly based on Refs.~\cite{Ostmeyer:2023azi,Luu:2026rts}.
	\end{abstract}

	\unitlength = 1em
	
	\newpage
	\tableofcontents
	
	\newpage
	\section{Introduction}
	
	The purpose of these notes is to collect state-of-the-art algorithms for the numerical simulation of fermionic systems.
	Most of the algorithms collected here have been published in Refs.~\cite{Ostmeyer:2023azi,Luu:2026rts} and implemented in \texttt{NSL}~\cite{NSL}.
	While some algorithms might be presented in a new and streamlined manner, none of them has been derived originally for these notes.
	
	Motivation as well as solid prior knowledge of the following topics are taken for granted:
	the Hubbard-Stratonovich transformation (see e.g.\ Ref.~\cite{ising} for a basic introduction) and, more generally, how to put fermions on a computer (see e.g.\ Ref.~\cite{Wynen:2018ryx}),
	basics of quantum Monte Carlo (QMC) simulations (see e.g.\ Refs.~\cite{ALF/SciPostPhysCodeb.1-v2.4,Ostmeyer:2025agg})
	as well as the analysis of Monte Carlo data in general (see e.g.\ Refs.~\cite{Wolff:2003sm,comp-avg})
	and, if required, Euclidean correlators in particular (see e.g.\ Refs.~\cite{Ostmeyer:2025igc,hadron}).
	
	These notes are structured as follows.
	In \cref{sec:sausage}, we introduce the ``sausage'' formalism as a way to reduce the dimensionality of the fermion matrix.
	Apart from briefly mentioning pseudo-fermions in \cref{sec:large_vol} as a way to deal with extremely large volumes, we will work in the ``sausage'' formalism throughout these notes.
	This implies that all algorithms deal with the exact fermion determinant.
	
	We provide some theoretical bounds on the computational complexity in \cref{sec:complexity_bounds} as a benchmark for our algorithms.
	The total runtime of different QMC algorithms is estimated in \cref{sec:update_schemes} and the measurement costs are discussed in \cref{sec:measurements}.

	We proceed with a general approach to calculate observables within this formalism in \cref{sec:observables}.
	The ``sausage'' formalism involves a large number of matrix products and the order in which the products are performed can make a great difference for the runtime.
	We explain the optimal strategies for the treatment of these matrix products in \cref{sec:product_accumulation}.
	
	\Cref{sec:naive_small_volume,sec:stable_small_volume,sec:med_vol_high_T,sec:med_vol_low_T,sec:large_vol,sec:large_volume_cold} provide a glossary-style collection of algorithms tailored specifically to a given regime.
	Depending on the spatial volume and the temperature, different algorithms provide the optimal trade off between stability and efficiency.
	The applicability and respective runtimes are summarised in \cref{tab:regimes_runtimes}.
	
	Finally, a more exotic approach applicable for very low fermion densities is explained in \cref{sec:low_filling}.\\
	
	\begin{table}[hb]
		\centering
		\begin{tabular}{c|c|c|c|}
			& $V\lesssim100$ & $20\lesssim V \lesssim 1000$ & $V\gtrsim 1000$ \\ \hline
			$\beta E \lesssim 20$ & sec.~\ref{sec:naive_small_volume}, $\ordnung{V^3 N_t}$ & sec.~\ref{sec:med_vol_high_T}, $\ordnung{V^2 N_t} + \ordnung{V^3}$ & sec.~\ref{sec:large_vol}, $\Omega\br{V N_t}$ \\ \hline
			$\beta E \gtrsim 20$ & sec.~\ref{sec:stable_small_volume}, $\ordnung{V^3 N_t}$ & sec.~\ref{sec:med_vol_low_T}, $\ordnung{V^2 N_t} + \ordnung{V^3\beta}$ & sec.~\ref{sec:large_volume_cold}, $\Omega\br{V N_t}$ \\ \hline
		\end{tabular}
		\caption{Summary of the different regimes in spatial volume $V$ and effective inverse temperature $\beta E$, where $E$ is some characteristic energy scale, together with the respective best suited algorithms and their runtimes. The regions denoted by `$\lesssim$' and `$\gtrsim$' are to be interpreted quite liberally. The runtime in terms of $V$ and the number of time slices $N_t$ is provided for simulations that require force calculations but no extensive observable measurements. $\Omega()$ means that the runtime is at least as high, probably (much) worse.}\label{tab:regimes_runtimes}
	\end{table}
	
	\newpage
	\section{The ``sausage''}
	\label{sec:sausage}

We start with a fermionic system of spatial extent $V$ and inverse temperature $\beta$.
The fermion dynamics are governed by the Hamiltonian
\begin{align}
	\mathcal{H}_\text{fermion} &= \sum_{ij} c_i^\dagger K_{ij} c_j^\pdagger\,,
\end{align}
where $c_i^\dagger$ and $c_j^\pdagger$ are creation and annihilation operators at the spatial positions $i,j$, respectively.
Any potential fields (including Hubbard-Stratonovich transformations) have already been included into the $V\times V$-dimensional hopping matrix $K$.
Throughout these notes we focus exclusively on this part of the (potentially much larger and more complicated) quantum system.

For lattice simulations, $\beta$ is discretised into $N_t$ (imaginary) time slices of extent $\delta\equiv\frac{\beta}{N_t}$.
Without derivation and going into detail about a particular physical model, such a system can be described by the real or complex vector fields $\varphi$ following the probability density
\begin{align}
	p(\varphi) &\propto \det M
\end{align}
governed by the determinant of the fermion matrix
\begin{align}
	M_{t',t} &= \delta_{t',t} - \delta_{t',t+1}M_t\,,\\
	M_t &= \eto{K_t}\,.
\end{align}
Anti-periodic boundary conditions in the time direction are implicitly assumed.
$K_t$ is a time-dependent version of the initial hopping matrix $\delta K$ with the time dependence included through the fields.
The factor $\delta$ has been absorbed into $K_t$, so each $K_t$ approaches zero with increasing $N_t$.
$M_t$ is also $V\times V$-dimensional, while the full $M$ is $\br{V N_t}\times \br{V N_t}$-dimensional.
In block-matrix notation $M$ takes the form
\begin{align}
	M &= \begin{pmatrix}
		1 & 0 & \cdots & 0 & M_{N_t-1} \\
		-M_0 & 1 & 0 & \cdots & 0 \\
		0 & -M_1 & 1 & \ddots & \vdots \\
		\vdots & \ddots & \ddots & \ddots & 0 \\
		0 & \cdots & 0 & -M_{N_t-2} & 1 
	\end{pmatrix}\,.
\end{align}

Now observe that
\begin{align}
	\hat M &\coloneqq 1 + \prod_t M_t\\
	\Rightarrow \det M &= \det\hat M\,,
\end{align}
which allows us to define the ``sausage'' $\hat M$ using the time-ordered product $\prod_t M_t\equiv M_0M_1\cdots M_{N_t-1}$.
The inverse of the sausage $\hat M^{-1}$ is often referred to as the Green's function~\cite{BSS:1981}.
Since the probability density mentioned above only depends on the determinant, we never need the full fermion matrix. Instead, we always replace it with the ``sausage'' $\hat M$.

Another important quantity is the Hamiltonian
\begin{align}
	H &= -\log \det \hat M + \cdots\,,\label{eq:hamiltonian_alg}
\end{align}
that might contain some bosonic contributions and additional fermionic flavours (i.e.\ multiple distinct fermion matrices) as well as canonical momenta.
Note that this artificial Hamiltonian used for Quantum Monte Carlo simulations already contains the factor $\beta$ (making it dimensionless) and is not to be confused with the physical Hamiltonian $\mathcal{H}$ of dimension energy.

\section{Bounds on computational complexity}\label{sec:complexity_bounds}

Without further knowledge of advanced algorithms, we can identify some lower and upper bounds on the computational complexity of fermionic QMC simulations.
For this we need to distinguish between the Monte Carlo updates and measurements as well as between runtime and memory requirements.
A full summary of the different complexity bounds is provided in \cref{tab:complexity_bounds}.

\begin{table}[ht]
	\centering
	\begin{tabular}{c|c|c|}
		& update & measurement \\ \hline
		runtime & $\Omega\br{VN_t}$ and $\ordnung{V^3N_t^3}$ & $\Omega\br{V^2N_t^2}$ and $\ordnung{V^3N_t^3}$ \\ \hline
		memory & $\Omega\br{VN_t}$ and $\ordnung{V^2N_t^2}$ & $\Theta\br{V^2N_t^2}$ \\ \hline
	\end{tabular}
	\caption{Summary of the computational complexities for runtime and memory requirements. An `update' is considered to be global in the sense that all the $\Theta\br{VN_t}$ field variables need to be updated at least once. A `measurement` includes all single-fermion observables, equivalent to the entire inverse $\br{V N_t}\times \br{V N_t}$-dimensional fermion matrix $M^{-1}$. $\Omega()$, $\ordnung{}$, and $\Theta()$ mean that the complexity is at least, at most, or exactly as high, respectively. The lower bounds are rigorous and not necessarily achievable. The upper bounds assume a na\"ive inversion of the full fermion matrix $M$.}\label{tab:complexity_bounds}
\end{table}

\subsection{Updates}

\label{sec:update_schemes}

To keep things tractable, we assume that there is a fixed number of real or complex fields with dimension $VN_t$ each.
Thus, any update of all field variables will involve at least $\Omega(VN_t)$ many operations.
This is also the minimal memory requirement.
On the other hand, the full (not ``sausage'' version) fermion matrix $M$ can be stored using $\ordnung{V^2N_t^2}$ memory and inverted in a runtime of order $\ordnung{V^3N_t^3}$.
Any algorithm using even more resources is clearly sub-optimal and should be avoided.

Let us briefly review how some of the most common algorithms compare to these theoretical bounds.

\subsubsection{The Hybrid Monte Carlo (HMC) algorithm}

The HMC algorithm~\cite{Duane1987} performs global updates of all the continuous field variables at once.
Therefore runtime and memory requirements of a single HMC step are equivalent to those for a full update from \cref{tab:complexity_bounds}.
It goes beyond the scope of these notes to dive into issues of minimal autocorrelation and other tunable properties of the algorithm.
We refer to Ref.~\cite{Ostmeyer:2025agg} for a hands-on introduction of the HMC and to Refs.~\cite{Brower:2011av,Wynen:2018ryx} for the derivation of the HMC formalism in a simple but non-trivial fermionic system.

An HMC step requires one evaluation of the Hamiltonian~\eqref{eq:hamiltonian_alg} (for the accept/reject step), the molecular dynamics (numerical integration of the equations of motion) and the calculation of fermionic forces $\del{H}{\varphi}$.
The Hamiltonian evaluation is necessary for any Monte Carlo update and a single evaluation per global update is the theoretical minimum.
The molecular dynamics by themselves scale as $\Theta(VN_t)$.
As a rule, the update runtime is dominated by the force calculation, a single-time observable measurement.

Thus, the lower bound on HMC complexity is (little surprising) $\Omega(VN_t)$.
There is no overhead complexity added by the algorithm, so (the fermionic contributions to) the true runtime and memory scalings are identical to those of Hamiltonian and force evaluations, see \cref{tab:regimes_runtimes}.

\subsubsection{The Blankenbecler-Scalapino-Sugar (BSS) algorithm}

The BSS algorithm~\cite{BSS:1981} (Sec.\ V) performs local updates, so one needs $\Theta(VN_t)$ BSS steps (a sweep) for a full update of all the field variables.
Each accepted local update comes with a runtime of $\Theta\br{V^2}$.
Rejected updates are $\ordnung{1}$, so the overall coefficient can be small.
On the other hand, many rejections come with long autocorrelation times.

Thus, the lower bound on BSS runtime complexity is $\Omega(V^3N_t)$.
This is a natural consequence of the intrinsic use of the ``sausage'' formalism by the BSS algorithm.
It comes with the additional overhead of an order $\Theta\br{V^3}$ (dense) or $\Theta\br{V^2}$ (sparse) update for every time slice, i.e.\ $\ordnung{V^3N_t}$ in total.

An important implication is that the BSS algorithm does not profit significantly from sparse matrix multiplications.
The lower bound is fixed to $\Omega(V^3N_t)$ by the necessity of a full sweep of local updates.

In contrast to the HMC, the BSS algorithm is applicable to discrete as well as continuous variables.

\subsection{Measurements}

\label{sec:measurements}

The computational complexity of measurements is highly observable-dependent.
Here, we provide the bounds for the calculation of all single-fermion observables.
This is equivalent to the determination of the full time-dependent fermionic Green's function.
In other words, we need the inverse of the $\br{V N_t}\times \br{V N_t}$-dimensional fermion matrix $M^{-1}$.
Since there are $V^2N_t^2$ elements to $M$ and $M^{-1}$, this is the lower bound for runtime and memory.
The na\"ive inversion requires $\ordnung{V^3N_t^3}$ steps, so that once more any algorithm with worse scaling should be avoided.

We remark that the primary bottleneck for measurements is the sheer number of results required.
An observable involving $s$ different times, will necessarily scale with $\Omega(N_t^s)$ (unless some measurements are dropped, e.g.\ for reasons of translational symmetry).
This is why the full Green's function $G(t_1,t_2)$ scales with $\Omega(N_t^2)$.
On the bright side, observables involving multiple fermions (e.g.\ two-particle correlators) can be calculated within the same computational complexity class as single fermion observables as long as they do not involve more different times.

Of course, some single-time observables (like the force) and those that are time-independent might be calculable in $\Omega(VN_t)$.

\section{Observables}

\label{sec:observables}

The vast majority of observables ever needed can be derived in the so-called response formalism relating the observable
\begin{align}
	\mathcal{A} &= - \del{H}{\alpha}
\end{align}
to some derivative of the Hamiltonian with appropriately chosen $\alpha$. For instance, the force needed for hybrid Monte Carlo (HMC) simulations is a response with respect to the main variable, e.g.\ the Hubbard field $\alpha=\phi$ for the Hubbard model.
If $\mathcal{A}$ is a physical observable (as opposed to an algorithmic one), we can obtain $\alpha$ using the relation
\begin{align}
	\mathcal{A} &= - \erwartung{\del{}{\alpha}\beta \mathcal{H}}\,.
\end{align}
A simple example is the total energy of the system for which we have to use $\alpha=\beta$.

It follows that all (relevant) fermionic observables can be written as a derivative of the fermion matrix
\begin{align}
	\mathcal{A} &= \del{\log\det\hat M}{\alpha}\\
	&= \frac{1}{\det\hat M} \del{\det \hat M}{\alpha}\\
	&= \frac{1}{\det\hat M}  \det\hat M \tr\br{\hat M^{-1} \,\del{\hat M}{\alpha}}\\
	&= \tr\br{\hat M^{-1}\sum_t\prod_{t'<t}M_{t'}\,\del{M_t}{\alpha}\prod_{t''>t}M_{t''}}\\
	&= \sum_t \tr\br{\prod_{t''>t}M_{t''}\,\hat M^{-1}\prod_{t'<t}M_{t'}\,\del{M_t}{\alpha}}\,,\label{eq:response_result}
\end{align}
where we used Jacobi's formula for the derivative of the determinant and subsequently expanded $\hat M$ using the product definition. For time-dependent observables the derivative only acts on a given time slice, say $t_0$, and the outer sum can be dropped
\begin{align}
	\mathcal{A}(t_0) &= \tr\br{\prod_{t''>t_0}M_{t''}\,\hat M^{-1}\prod_{t'<t_0}M_{t'}\,\del{M_{t_0}}{\alpha}}\,.
\end{align}

\subsection{Shortcut for simple operators}

\label{sec:shortcut}

In a sloppy abuse of notation we can relate $K_t=\delta \mathcal{H}$ and therefore for simple enough physical operators
\begin{align}
	\del{M_t}{\alpha} &\approx M_t \del{K_t}{\alpha}\\
	&= M_t \frac{1}{N_t} \hat{\mathcal{A}}\,.
\end{align}
where $\hat{\mathcal{A}}$ denotes the operator corresponding to the observable $\mathcal{A}$.
Inserting this result into equation~\eqref{eq:response_result}, we obtain
\begin{align}
	\mathcal{A} &\approx \frac{1}{N_t} \sum_t \tr\br{\prod_{t''>t}M_{t''}\,\hat M^{-1}\prod_{t'\le t}M_{t'}\,\hat{\mathcal{A}}}\,.\label{eq:response_simple}
\end{align}
Thus, the expectation value of a fermionic observable can often be obtained by plugging the corresponding operator into equation~\eqref{eq:response_simple}.
More rigorously, we take the thermal trace of the operator multiplied by the Green's function at every discrete time.

A simple example for which this shortcut works very well is the fermion number $\mathcal{A}=n$, $\hat{\mathcal{A}}=\frac 1V \mathds{1}$ (a response to the chemical potential $\alpha=\mu$), which reduces to
\begin{align}
	n &= \frac{1}{VN_t} \sum_t \tr\br{\prod_{t''>t}M_{t''}\,\hat M^{-1}\prod_{t'\le t}M_{t'}}\\
	&= \frac 1V \tr\br{\hat M^{-1}\prod_{t'}M_{t'}}\\
	&= 1 - \frac 1V \tr \hat M^{-1}\,.\label{eq:fermion_number}
\end{align}

\subsection{Matrix element derivatives}

Another important class of observables (e.g.\ the single particle correlator) is described by the response with respect to some specific matrix element $M_{t_0x,ty}$ of the full fermion matrix.
By Jacobi's formula this response is, in fact, identical to the corresponding matrix element of the full time-dependent Green's function, that is inverse fermion matrix.
The corresponding derivative
\begin{align}
	\br{M^{-1}}_{ty,t_0x} &= \del{\log\det\hat M}{M_{t_0x,ty}}\\
	&=\begin{cases}
		\tr\br{\prod\limits_{t'' = t_0-1}^0 M_{t''}^{-1}\,\hat M^{-1}\prod\limits_{t'<t}M_{t'}\,\delta_{y}^x} & \text{if } t_0\le t\,,\\
		-\tr\br{\prod\limits_{t'' \ge t_0}M_{t''}\,\hat M^{-1}\prod\limits_{t'<t}M_{t'}\,\delta_{y}^x} & \text{else}
	\end{cases}\\
	&=\begin{cases}
		\brr{\br{\prod\limits_{t'=t-1}^{t_0}M_{t'}^{-1} + \prod\limits_{t'=t}^{N_t-1}M_{t'}\prod\limits_{t''=0}^{t_0-1}M_{t''}}^{-1}}_{yx} & \text{if } t_0\le t\,,\\
		-\brr{\br{\prod\limits_{t''=t-1}^{0}M_{t''}^{-1}\prod\limits_{t'=N_t-1}^{t_0}M_{t'}^{-1} + \prod\limits_{t'=t}^{t_0-1}M_t'}^{-1}}_{yx} & \text{else}
	\end{cases}\label{eq:sausage_all_in_denominator}
\end{align}
can also directly be expressed in the ``sausage'' formalism. Here $\delta^x_y$ denotes a $V\times V$-matrix that has a single unit entry at the $(x,y)$ position and is zero everywhere else. Note that in the special case of $t_0=t$ the term $\prod_{t'=t-1}^{t_0}M_{t'}^{-1}=\mathds{1}$ becomes an empty product because $t'\ge t_0$ is decreasing.

From this relation it also becomes clear why the inverse fermion matrix is sufficient for all fermionic observables that can be expressed in the response formalism.
Applying the chain rule, we can rewrite any derivative as
\begin{align}
	\mathcal{A} &= \del{\log\det\hat M}{\alpha}\\
	&= \sum_{ij} \del{\log\det\hat M}{M_{ij}}\, \del{M_{ij}}{\alpha}\\
	&= \sum_{ij} \br{M^{-1}}_{ji}\, \del{M_{ij}}{\alpha}\\
	&= \tr\br{M^{-1}\, \del{M}{\alpha}}\,,
\end{align}
that is a linear combination of simple derivatives $\del{M_{ij}}{\alpha}$ with prefactors from $M^{-1}$.

\subsection{Discrete time artifacts}

To leading order in $\delta$, the expression can be further simplified to 
\begin{align}
	\mathcal{A} &= \sum_t \tr\br{\prod_{t''>t}M_{t''}\,\hat M^{-1}\prod_{t'\le t}M_{t'}\,\del{K_t}{\alpha}} + \ordnung{\delta}\,,
\end{align}
which is typically sufficient and used in most calculations. $\ordnung{\delta}$-improved approximations are readily available
\begin{align}
	\mathcal{A} &= \frac12 \sum_t \tr\br{\prod_{t''>t}M_{t''}\,\hat M^{-1}\prod_{t'\le t}M_{t'}\,\del{K_t}{\alpha}+\prod_{t''\ge t}M_{t''}\,\hat M^{-1}\prod_{t'< t}M_{t'}\,\del{K_t}{\alpha}} + \ordnung{\delta^2}\\
	&= \sum_t \tr\br{\prod_{t''\gtrsim t}M_{t''}\,\hat M^{-1}\prod_{t'\lesssim t}M_{t'}\,\del{K_t}{\alpha}} + \ordnung{\delta^2}\,,\\
	\prod_{t'\lesssim t}M_{t'} &\coloneqq \prod_{t'< t}M_{t'} \, M_t^{1/2}\,,\\
	M_t^{1/2} &= \eto{\frac12 K_t}\,.
\end{align}
Both formulae are second order approximations of the integral
\begin{align}
	\mathcal{A} &= \int_{0}^{1}\md x\sum_t \tr\br{M_t^{1-x}\prod_{t''> t}M_{t''}\,\hat M^{-1}\prod_{t'< t}M_{t'}\,M_t^x\,\del{K_t}{\alpha}}\,.
\end{align}
However, higher than second order approximations would exceed the order of the Trotterization in the initial formulation of the fermion matrix and are therefore irrelevant. We only remark that the exact derivative is calculable and should in fact be evaluated if the exponential $M_t=\eto{K_t}$ is already truncated to some finite polynomial. In that case the derivative of the polynomial is to be used instead of the approximations listed above.

\subsection{Multiple operators}

A similar derivation allows to write a time-dependent product of operators as
\begin{multline}
	\mathcal{A}_1(t_1)\mathcal{A}_2(t_2)\cdots \mathcal{A}_s(t_s) =\\
	 \tr\br{\prod_{t_s''\gtrsim t_s}M_{t_s''}\,\hat M^{-1}\prod_{t_1'\lesssim t_1}M_{t_1'}\,\del{K_{t_1}}{\alpha_1}\br{\prod_{t_1''\lesssim t_1}M_{t_1''}}^{-1}\hat M^{-1}\prod_{t_2'\lesssim t_2}M_{t_2'}\,\del{K_{t_2}}{\alpha_2}\br{\prod_{t_2''\lesssim t_2}M_{t_2''}}^{-1}\cdots M^{-1}\prod_{t_s'\lesssim t_s}M_{t_s'}\,\del{K_{t_s}}{\alpha_s}}\\
	 \quad + \ordnung{\delta^2}\,,\label{eq:operator_product_long}
\end{multline}
or, more readably,
\begin{align}
	\mathcal{A}_1(t_1)\mathcal{A}_2(t_2)\cdots \mathcal{A}_s(t_s) &=\tr\br{\br{1-\hat M^{-1}}\del{K_{t_1}}{\alpha_1}(t_1)\del{K_{t_2}}{\alpha_2}(t_2)\cdots\del{K_{t_s}}{\alpha_s}(t_s)} + \ordnung{\delta^\nu}\,,\label{eq:operator_product_short}
\end{align}
where the exponent $\nu$ depends on the respective approximation of the matrix derivative, $1-\hat M^{-1}$ acts like the Fermi function, and
\begin{align}
	\del{K_{t_i}}{\alpha_i}(t_i) &\coloneqq \prod_{t'\le t_i}M_{t_i'}\,\del{K_{t_i}}{\alpha_i}\br{\prod_{t''\le t_i}M_{t_i''}}^{-1}\label{eq:heisenberg_pic_prod}\\
	&= \eto{K_0}\eto{K_1}\cdots\eto{K_{t_i}}\,\del{K_{t_i}}{\alpha_i}\,\eto{-K_{t_i}}\cdots\eto{-K_1}\eto{-K_0}\label{eq:heisenberg_picture}
\end{align}
can be interpreted as the operator $\hat{\mathcal{A}}_i(t_i)\sim \del{K_{t_i}}{\alpha_i}(t_i)$ in the Heisenberg picture. While the latter formulation~\eqref{eq:operator_product_short} is easier to read, the former version~\eqref{eq:operator_product_long} is typically more efficient to implement (after re-writing the inverses as in eq.~\eqref{eq:heisenberg_picture}). Of course, the `$\le$' in equation~\eqref{eq:heisenberg_pic_prod} and the `$\lesssim$' in equation~\eqref{eq:operator_product_long} can be replaced as desired.

Without going into any further detail, these expressions show that any fermionic quantity in the ``sausage''-formalism can be calculated with the knowledge of $\det \hat M$ and all possible combinations of $\prod_{t'>t_1}M_{t'}\, \hat M^{-1} \prod_{t''<t_2}M_{t''}$ (allowing negative times $t_{1,2}$ and inverses $M_{t'}^{-1}$). Therefore, an implementation only has to compute said quantities. In the following, we will explore efficient and stable ways to implement them. Depending on the regime, the trade off between efficiency and stability favours some algorithm or another. See \cref{tab:regimes_runtimes} for a summary.

\section{Product accumulation tricks}

\label{sec:product_accumulation}

Usually, the runtime will be dominated by the products $\prod_{t'>t_1}M_{t'}\, \hat M^{-1} \prod_{t''<t_2}M_{t''}$ since computing them for all possible choices of $t_{1,2}$ naively scales as $\ordnung{V^3N_t^3}$, or as $\ordnung{V^3N_t^2}$ if $t_1=t_2$ (e.g.\ for force calculations). It is therefore crucial for the runtime to introduce one of the following tricks.

\subsection{Dense pre- and suffix calculations}\label{sec:prod_trick_dense}

\begin{enumerate}
	\item Multiply and store $\prod_{t'>t}M_{t'}$ and $\prod_{t'<t}M_{t'}$ for each $t$
	\item Combine $\prod_{t'>t_1}M_{t'}\, \hat M^{-1} \prod_{t''<t_2}M_{t''}$ as required
\end{enumerate}

Thus, the runtime reduces to $\ordnung{V^3N_t^2}$, or $\ordnung{V^3N_t}$ if $t_1=t_2$.
Memory scales as $\ordnung{V^2N_t}$ and is unlikely to be the bottleneck.

\subsection{Sparse pre- and suffix calculations}\label{sec:prod_trick_sparse_tr}

The motivation for the following modified procedure is that multiplication with a single time slice $M_t$ can be cheaper than $\ordnung{V^3}$. Using sparse matrices, it can typically be reduced to $\ordnung{V^2}$. The 2nd step in \cref{sec:prod_trick_dense} necessarily multiplies dense matrices and therefore does not profit from the sparsity. In contrast, the procedure below requires only (potentially sparse) multiplications with $M_t$.

The following method is only efficient if one is interested in a trace of the form
\begin{align}
	\mathcal{A}(t_1,t_2) &= \tr\brr{\prod_{t'>t_1}M_{t'}\, \hat M^{-1} \prod_{t''<t_2}M_{t''}}.
\end{align}
For a (significantly more involved) method to obtain the entire matrix $\prod_{t'>t_1}M_{t'}\, \hat M^{-1} \prod_{t''<t_2}M_{t''}$ efficiently, see \cref{sec:prod_trick_sparse}.

\begin{enumerate}
	\item Multiply and store $\prod_{t'>t}M_{t'} \, \hat M^{-1}$ and $\prod_{t'<t}M_{t'}$ for each $t$
	\item Combine $\mathcal{A}(t_1,t_2) = \sum_{ij}\brr{\prod_{t'>t_1}M_{t'}\, \hat M^{-1}}_{ij} \brr{\prod_{t''<t_2}M_{t''}}_{ji}$ as required
\end{enumerate}

Thus, for sparse matrices $M_t$ the runtime reduces to $\ordnung{V^2N_t^2}$, or $\ordnung{V^2N_t}$ if $t_1=t_2$.
Memory scales as $\ordnung{V^2N_t}$.

\subsection{Stabilised dense pre- and suffix calculations}\label{sec:prod_trick_dense_stab}

The formulation~\eqref{eq:sausage_all_in_denominator} of $M^{-1}$ is numerically more stable to compute than the slightly simpler version used in \cref{sec:prod_trick_dense}.
The overhead in runtime is just a small constant factor.
Thus, we obtain the following stabilised prescription.

\begin{enumerate}
	\item Multiply and store $\prod_{t'=t-1}^{0}M_{t''}^{-1}$ and $\prod_{t'=N_t-1}^{t}M_{t'}^{-1}$ for each $t$
	\item For each $t_2$ multiply and store all relevant $\prod_{t=t_2}^{t_1-1}M_t$
	\item Combine $\prod_{t''=t_2-1}^{0}M_{t''}^{-1}\prod_{t'=N_t-1}^{t_1}M_{t'}^{-1}$ as required
	\item Use stabilised procedure from \cref{sec:stable_small_volume} for the inversion~\eqref{eq:sausage_all_in_denominator}
\end{enumerate}

Thus, the runtime reduces to $\ordnung{V^3N_t^2}$, or $\ordnung{V^3N_t}$ if $t_1=t_2$.
Memory scales as $\ordnung{V^2N_t}$ and is unlikely to be the bottleneck.

\subsection{Sparse recursive accumulation}\label{sec:prod_trick_sparse}

As in \cref{sec:prod_trick_sparse_tr}, the following procedure allows to exploit the sparse nature of $M_t$ when applicable.
It should be used if the full matrices $\prod_{t'>t_1}M_{t'}\, \hat M^{-1} \prod_{t''<t_2}M_{t''}$ are required rather than only traces thereof.

\begin{enumerate}
	\setcounter{enumi}{-1}
	\item Start with $\hat M^{-1}$
	\item \begin{enumerate}
		\item If $t_1=t_2$, multiply $$\prod\limits_{t'>\left\lceil \nicefrac{N_t}{2}\right\rceil}\!\!\! M_{t'} \,\hat M^{-1}\,,\;\hat M^{-1} \prod\limits_{t''<\left\lfloor \nicefrac{N_t}{2}\right\rfloor}\!\!\!M_{t''}$$
		\item Else, accumulate $$\hat M^{-1}\,,\;\prod\limits_{t'>\left\lceil \nicefrac{N_t}{2}\right\rceil}\!\!\! M_{t'} \,\hat M^{-1}\,,\;\hat M^{-1} \prod\limits_{t''<\left\lfloor \nicefrac{N_t}{2}\right\rfloor}\!\!\!M_{t''}\,,\;\prod\limits_{t'>\left\lceil \nicefrac{N_t}{2}\right\rceil}\!\!\! M_{t'} \,\hat M^{-1} \prod\limits_{t''<\left\lfloor \nicefrac{N_t}{2}\right\rfloor}\!\!\!M_{t''}$$
	\end{enumerate}
	\item Construct
	\begin{enumerate}
		\item If $t_1=t_2$, $$\prod\limits_{t'>\left\lceil \nicefrac{N_t}{4}\right\rceil}\!\!\! M_{t'}\, \hat M^{-1}\,,\; \prod\limits_{t'>\left\lceil \nicefrac{N_t}{2}\right\rceil}\!\!\! M_{t'} \, \hat M^{-1} \prod\limits_{t''<\left\lfloor \nicefrac{N_t}{4}\right\rfloor}\!\!\!M_{t''}\,,\; \prod\limits_{t'>\left\lceil \nicefrac{3}{4}N_t\right\rceil}\!\!\! M_{t'} \, \hat M^{-1} \prod\limits_{t''<\left\lfloor \nicefrac{N_t}{2}\right\rfloor}\!\!\!M_{t''}\,,\; \hat M^{-1} \prod\limits_{t''<\left\lfloor \nicefrac{3}{4}N_t\right\rfloor}\!\!\!M_{t''}$$
		\item Else, for all matrices $Q$ from step 1 $$Q\,,\;\prod\limits_{t'>\left\lceil \nicefrac{3}{4}N_t\right\rceil}\!\!\! M_{t'} \, Q\,,\; Q \prod\limits_{t''<\left\lfloor \nicefrac{N_t}{4}\right\rfloor}\!\!\!M_{t''}\,,\;\prod\limits_{t'>\left\lceil \nicefrac{3}{4}N_t\right\rceil}\!\!\! M_{t'} \, Q \prod\limits_{t''<\left\lfloor \nicefrac{N_t}{4}\right\rfloor}\!\!\!M_{t''}$$
	\end{enumerate}
	\item[$\vdots$]
	\item[$n$.] Iterate
	\begin{enumerate}
		\item If $t_1=t_2$, construct $2^n$ matrices from $2^{n-1}$ in previous step, multiplying $\left\lfloor\frac{N_t}{2^n}\right\rfloor$ time slices from left and right.
		\item Else, construct $4^n$ matrices from $4^{n-1}$ in previous step, multiplying $\left\lfloor\frac{N_t}{2^n}\right\rfloor$ time slices from left, right, neither, and both.
	\end{enumerate}
\end{enumerate}

This procedure converges after $\left\lceil\log_2 N_t\right\rceil$ steps.
For sparse matrices $M_t$ the runtime reduces to $\ordnung{V^2 N_t^2}$, or $\ordnung{V^2 N_t\log N_t}$ if $t_1=t_2$.
Memory scales as $\ordnung{V^2N_t^2}$ for all combinations of $t_{1,2}$, or as $\ordnung{V^2 N_t}$ if $t_1=t_2$, and might be relevant in the former case.

\section{Small volume $V\lesssim 20$, high temperature (small $\beta\lesssim 20$)}\label{sec:naive_small_volume}

In this easiest possible regime the implementation can proceed without elaborate tricks. The small volume guarantees that dense matrix operations are computationally cheap and at small\footnote{Keep in mind that $\beta$ itself is not a dimensionless quantity, so the smallness is truly about the product $\beta E\lesssim 20$ with some characteristic energy scale $E$.} $\beta$ no numerical instabilities are to be expected. Proceed as follows:

\begin{enumerate}
	\item Diagonalise $K_t = U_t D_t U_t^\dagger$
	\item Contract $M_t = U_t \eto{D_t} U_t^\dagger$ \label{item:def_M_t}
	\item Multiply and diagonalise $\prod_t M_t = \hat S D' \hat S^{-1}$ (the product is not hermitian in general) \label{item:diagonalise_prod}
	\item Set $\hat D = 1+D'$ \label{item:diagonal_sum}
	\item Sum $\log\det\hat M = \tr\log\hat D$ \label{item:log_det_M}
	\item Contract $\hat M^{-1} = \hat S \hat D^{-1} \hat S^{-1}$ \label{item:get_M_inv}
	\item Use sec.~\ref{sec:prod_trick_dense} to multiply $\prod_{t'>t_1}M_{t'}\, \hat M^{-1} \prod_{t''<t_2}M_{t''}$ as required
\end{enumerate}

The runtime of each dense matrix operation (diagonalisation and matrix multiplication) scales as $\ordnung{V^3}$ and there are at least $\ordnung{N_t}$ such operations. Thus, the runtime scales at least as $\ordnung{V^3 N_t}$. If various combinations of $t_{1,2}$ are required, the runtime increases to $\ordnung{V^3 N_t^2}$, see sec.~\ref{sec:prod_trick_dense}.

The main advantage of this algorithm is that it is easy to implement. In addition, it requires only about half as many dense matrix operations as the stabilised method of \cref{sec:stable_small_volume}, though it remains in the same computational complexity class. In terms of accuracy and numerical stability it is at best equivalent to the stabilised method.

\section{Small volume $V\lesssim 20$, any temperature (any $\beta$)}\label{sec:stable_small_volume}

In this regime the entire focus lies on maximising numerical stability that might be jeopardised by the large values of $\beta$ and, in consequence, exponentially large eigenvalues of $\hat M$.
Since the spatial volume $V$ is small, runtime for dense matrix manipulations is uncritical.
The stabilised procedure has been derived in Ref.~\cite{Luu:2026rts} and relies heavily on a framework developed in Ref.~\cite{Bauer:2020stable}.

Conceptually, all steps proceed exactly equivalently to those in \cref{sec:naive_small_volume}. The difference is that matrices are not stored in the standard $V\times V$ representation, but as a product of three different matrices that are not contracted. The central matrix in this product is diagonal with entries sorted by magnitude, the outer matrices are basis transformations with entries of order unity.

Write any matrix in the form
\begin{align}
	A &= X D Y^{-1}\,,\label{eq:def_matrix_as_product}
\end{align}
where $D$ is real, diagonal, and sorted by absolute values of the entries. Moreover, $X$ and $Y$ have entries of order unity and $\left|\det X\right|=\left|\det Y\right|=1$. $A$ is never to be computed in this formalism, instead $X$, $Y$, and $D$ are stored independently. Now a matrix multiplication proceeds as follows
\begin{align}
	A_1 A_2 &= X_1 \brr{D_1 \br{Y_1^{-1} X_2} D_2} Y_2^{-1}\label{eq:prod_2_mat_stab}\\
	&= \brr{X_1 X'} D \brr{Y_2 Y'}^{-1}\\
	&= X D Y^{-1}\,,\label{eq:res_2_mat_stab}
\end{align}
where square brackets $\brr{\cdot}$ signify matrices to be contracted and round brackets $\br{\cdot}$ specify the order. In the intermediate step
\begin{align}
	\brr{D_1 \br{Y_1^{-1} X_2} D_2} &= X' D {Y'}^{-1}
\end{align}
the contracted matrix on the LHS is decomposed into the three matrices on the RHS using a QR decomposition ($Q$ is unitary and $R$ upper triangular). Then $X'=Q$, $D=\diag R$, and ${Y'}^{-1}=D^{-1}R$.
Note that the inversions $Q^{-1}=Q^\dagger$ and $R^{-1}$ are numerically very stable and cheap in $\ordnung{V^2}$.
Moreover, unitary and upper triangular matrices are closed under multiplication so that $X$ and $Y$ can be chosen unitary and upper triangular for all involved matrices.

Increased numerical stability can thus be achieved by replacing the steps~\ref{item:diagonalise_prod} through~\ref{item:log_det_M} as follows:

\begin{enumerate}
	\item Diagonalise $K_t = U_t D_t U_t^\dagger$
	\item Identify $M_t = U_t \eto{D_t} U_t^\dagger$ as in eq.~\eqref{eq:def_matrix_as_product} without contracting
	\item Use the stabilised procedure of eqs.~(\ref{eq:prod_2_mat_stab}-\ref{eq:res_2_mat_stab}) to multiply $\prod_t M_t = \tilde X \tilde D \tilde Y^{-1}$\\ (s.t.\ $\left|\det \tilde X\right|=\left|\det \tilde Y\right|=1$ and $\tilde D$ diagonal)
	\item QR-decompose $1+\tilde D\tilde Y^{-1}\tilde X = \hat Q \hat D \hat R$ (with $\diag \hat R = 1$),
	thus $\hat M = \tilde X \hat Q \hat D \hat R \tilde X^{-1}$
	\item Sum $\log\det\hat M = \imag\log\det\br{\hat Q} + \tr\log\hat D$ \label{item:log_det_M_stab}
	\item Contract $\hat M^{-1} = \br{\tilde X \hat R^{-1}} \hat D^{-1} \br{\hat Q^\dagger \tilde X^{-1}}$
\end{enumerate}

In step~\ref{item:log_det_M_stab} we used the prior knowledge that $\abs{\det({\hat Q})}=1$ and thus any contribution to the logarithm must be purely imaginary up to numerical precision.

Additional care has to be taken in the calculation of observables involving Green's function matrix elements $\br{M^{-1}}_{t_2,t_1}$ because products of the form $\prod_{t'>t_1}M_{t'}\, \hat M^{-1} \prod_{t''<t_2}M_{t''}$ turn out to be notoriously unstable even when using the stabilisation described above.
Instead, we resort to the form~\eqref{eq:sausage_all_in_denominator} and use the pre- and suffix trick from \cref{sec:prod_trick_dense_stab}.
This way, a separate inversion is performed for every arrangement of time slices.

\begin{enumerate}
	\setcounter{enumi}{6}
	\item Combine $\prod_{t=t_2}^{t_1-1}M_t=X' D' {Y'}^{-1}$ and $\prod_{t''=t_2-1}^{0}M_{t''}^{-1}\prod_{t'=N_t-1}^{t_1}M_{t'}^{-1} = Y'' {D''}^{-1} {X''}^{-1}$ as required for eq.~\eqref{eq:sausage_all_in_denominator} following sec.~\ref{sec:prod_trick_dense_stab}
	\item QR-decompose $D' \br{{Y'}^{-1} X''} + \br{{X'}^{-1}Y''} {D''}^{-1} = Q D R$
	\item Contract $\br{M^{-1}}_{t_2,t_1} = \br{X'' R^{-1}} D^{-1} \br{Q^\dagger {X'}^{-1}}$
\end{enumerate}

\section{Intermediate volume $20\lesssim V\lesssim 1000$, high temperature (small $\beta\lesssim 20$)}\label{sec:med_vol_high_T}

In this numerically stable regime the entire focus lies on improved runtime. At the cost of minimal accuracy loss, most matrix manipulations can be reduced to sparse operations, reducing the runtime of matrix-vector multiplications from $\ordnung{V^2}$ to $\ordnung{V}$ and matrix-matrix multiplications from $\ordnung{V^3}$ to $\ordnung{V^2}$. For this we rewrite
\begin{align}
	M_t &= \eto{K_t}\\
	&= \prod_{k=1}^n\br{1 + \gamma_k^{(n)} K_t} + \ordnung{\delta^{n+1}}\,,
\end{align}
where the (complex-valued) coefficients $\gamma_k^{(n)}$ have to be chosen so that the product equals the Taylor series of the exponential to the same order $n$. Typically, $n=4$ is a good compromise between accuracy and computational cost. The product is numerically more stable and computationally more efficient than the sum naively defining the Taylor series. For details see Ref.~\cite{Ostmeyer:2023xju}.

$K_t$ is never to be written out completely in matrix form, but always to be applied in a matrix-free formulation. Presume $K_t$ has $q$ non-zero entries per row (typically the number of nearest neighbours, or similar). Then the runtime of a matrix-vector multiplication $M_t\cdot v$ scales as $\ordnung{nqV}$ which is why the sparse formalism is only beneficial when $V>nq$.
Using $n=4$ and estimating $q\approx 5$, a typical pivoting volume lies around $V_p\approx 20$.

\subsection{Fastest sparse algorithm, $\beta\lesssim 10$}

The LU-decomposition is the fastest way to invert a general dense matrix, obtaining the determinant as a by-product.

\begin{enumerate}
	\item Start with $P_0=1$
	\item For $t=0,\dots,N_t-1$ apply $P_{t+1}= M_t \cdot P_t$
	\item LU-decompose $\hat M = 1+P_{N_t} = L U$ (with $\diag L=1$)
	\item Set $\hat D = 1+\diag U$
	\item Sum $\log\det\hat M = \tr\log\hat D$
	\item Invert and contract $\hat M^{-1} = U^{-1} L^{-1}$
	\item Use sec.~\ref{sec:prod_trick_sparse_tr} or sec.~\ref{sec:prod_trick_sparse} to multiply $\prod_{t'>t_1}M_{t'}\, \hat M^{-1} \prod_{t''<t_2}M_{t''}$ as required
\end{enumerate}

\subsection{Stabilised sparse algorithm, $\beta\gtrsim 10$}\label{sec:med_vol_high_T_stab}

The full diagonalisation is slower than the LU-decomposition, though still in the same computational complexity class $\ordnung{V^3}$. However, it substantially stabilises determinant and inverse calculations.

\begin{enumerate}
	\item Start with $P_0=1$
	\item For $t=0,\dots,N_t-1$ apply $P_{t+1}= M_t \cdot P_t$ \label{item:apply_prod_M}
	\item Diagonalise $\prod_t M_t = P_{N_t} = \hat S D' \hat S^{-1}$ (the product is not hermitian in general) \label{item:diag_prod_M}
	\item Set $\hat D = 1+D'$
	\item Sum $\log\det\hat M = \tr\log\hat D$
	\item Contract $\hat M^{-1} = \hat S \hat D^{-1} \hat S^{-1}$
	\item Use sec.~\ref{sec:prod_trick_sparse_tr} or sec.~\ref{sec:prod_trick_sparse} to multiply $\prod_{t'>t_1}M_{t'}\, \hat M^{-1} \prod_{t''<t_2}M_{t''}$ as required
\end{enumerate}

Both algorithms have a total runtime scaling at least as $\ordnung{nqV^2 N_t} + \ordnung{V^3}$. If various combinations of $t_{1,2}$ are required, the runtime increases to $\ordnung{nqV^2 N_t^2} + \ordnung{V^3}$, or to $\ordnung{nqV^2 N_t \log N_t} + \ordnung{V^3}$ if $t_1=t_2$, see sec.~\ref{sec:prod_trick_sparse}.

\section{Intermediate volume $20\lesssim V\lesssim 1000$, low temperature (large $\beta\gtrsim 20$)}\label{sec:med_vol_low_T}

In this regime both numerical stability and runtime have to be taken into consideration. 
Overall, the algorithm advocated here is a further stabilised and slightly slower version of the one presented in \cref{sec:med_vol_high_T_stab}. It has not been tested in practice yet.
We propose to use the algorithm in \cref{sec:stable_small_volume} with two modifications. First of all, the multiplications by $M_t$ are performed in the sparse formalism. In addition, the accumulating matrix $P_t=Q_tD_tR_t$ is not QR-decomposed at every step but only after $\nicefrac{\beta_0}{\delta}$ steps where $\beta_0\simeq 10$ is some reasonably large but numerically stable inverse temperature. These sporadic stabilisations should suffice to restore stability to the entire algorithm. The overhead $\ordnung{V^3N_t\frac{\delta}{\beta_0}}=\ordnung{V^3\frac{\beta}{\beta_0}}$ should not be too prohibitive compared to the main contribution $\ordnung{nqV^2 N_t}$.

\section{Large volume $V\gtrsim 1000$, high temperature (small $\beta\lesssim 20$), no sign problem}\label{sec:large_vol}

In this regime, any dense treatment of the ``sausage'' becomes prohibitively expensive and it might be necessary to switch to a matrix-free approach like the use of pseudo-fermions~\cite{Kennedy:2006ax,acceleratingHMC}. At this point, it is not completely clear whether keeping the ``sausage'' or reverting to the full fermion matrix is more efficient and it might well depend on other parameters like the temperature. We will not dive into detail here.

In short pseudo-fermions can be introduced as an additional auxiliary field.
Typically they are used for an even number of fermion flavours with unbroken particle-hole symmetry that can be combined to positive definite pairs. Then the fermion determinant can be rewritten using complex random variables $\chi$
\begin{align}
	\det\br{\hat M\hat M^\dagger} &\propto \int \mathcal{D}\chi^*\mathcal{D}\chi \,\eto{-\chi^\dagger \br{\hat M\hat M^\dagger}^{-1} \chi}\,,
\end{align}
where $\mathcal{D}\chi$ denotes the path integral over all dimensions of $\chi$. Thus, only matrix-vector multiplications and solves are required.

In principle, one can also rewrite the fermion determinant for individual fermion flavours, as long as the determinant is greater or equal to zero, i.e.\ there is no sign problem. This is rarely done in practice and added here more for the sake of completeness than for practical use. A real pseudo-fermionic field $\eta$ suffices in this case
\begin{align}
	\left|\det \hat M\right| &= \sqrt{\det\br{\hat M\hat M^\dagger}}\\
	 &\propto \int \mathcal{D}\eta \,\eto{-\frac12 \eta^\trans \br{\hat M\hat M^\dagger}^{-1} \eta}\,.
\end{align}

\section{Large volume $V\gtrsim 1000$, low temperature (large $\beta\gtrsim 20$) or sign problem (even mild)}

\label{sec:large_volume_cold}

Well, that's tough.

\section{Very low filling $\erwartung{n}\ll1$}

\label{sec:low_filling}

As a rule, fermionic simulations become much more complicated once a non-zero chemical potential $\mu$ is introduced.
The chemical potential shifts the system away from half filling, typically introducing a fermionic sign problem.
This problem is known to be NP-hard~\cite{Troyer:2004ge} and we will not elaborate on the great multitude of approaches in these notes.

In some cases, however, the addition of a chemical potential can simplify the simulations.
This happens when the filling (i.e.\ the average fermion number $\erwartung{n}$) is so low that the fermionic interaction becomes negligible.
As a consequence, the fermions can be treated as a perturbation to a free theory.

To see this, we write the suppression by the chemical potential explicitly in the ``sausage''
\begin{align}
	\hat M &= 1 + \eto{-\beta\mu}\prod_t M_t\,,
\end{align}
so that with equation~\eqref{eq:fermion_number} immediately follows
\begin{align}
	\hat M^{-1} &= 1 - \eto{-\beta\mu}\prod_t M_t + \ordnung{\eto{-2\beta\mu}}\\
	\Rightarrow n &= \frac1V \eto{-\beta\mu}\tr\prod_t M_t + \ordnung{\eto{-2\beta\mu}}
\end{align}
and therefore
\begin{align}
	\log\det\br{1+\eto{-\beta\mu}\prod_t M_t} &= \eto{-\beta\mu}\tr\prod_t M_t + \ordnung{\eto{-2\beta\mu}}\\
	&= V n + \ordnung{\eto{-2\beta\mu}}\,.
\end{align}

The following two methods are summarised from Ref.~\cite{Ostmeyer:2023azi} wherein more details can be found (especially in the supplementary material).

\subsection{Zero particles regime}

Neglecting all $\ordnung{\eto{-\beta \mu}}$-terms, fully quenched simulations can be performed.
Reweighting can be added if necessary.

\subsection{Single particle regime}

A particularly powerful method can be used at low filling for the calculation of observables that are normalised by the particle number.
Simulations can be performed in the single particle regime where the probability distribution is rescaled by $n$.
The Hamiltonian after rescaling reads
\begin{align}
	H_n &= H - \log n \label{eq:n_reweighted_hamiltonian}\\
	&= S_B - \log\tr\prod_t M_t + \ordnung{\eto{-\beta\mu}} + \text{const.}\,,\label{eq:single_particle_hamiltonian}
\end{align}
where $S_B$ is the bosonic action.

Subsequent reweighting simplifies the ratio of interest.
The average $\erwartung{\cdot}_n$ obtained this way allows to cancel correlations before averaging
\begin{align}
	\frac{\erwartung{\mathcal{A}}}{\erwartung{n}} &= \erwartung{\frac{\mathcal{A}}{n}}_n\,,
\end{align}
typically increasing precision considerably.

	\section*{Acknowledgements}
	Many thanks to Evan Berkowitz, Pavel Buividovich, Tom Luu, Petar Sinilkov and Finn Temmen for highly insightful discussions.
	The author would also like to thank his shower for providing such an inspiring environment.
	This work was funded in part by the Deutsche Forschungsgemeinschaft (DFG, German Research Foundation) as part of the CRC 1639 NuMeriQS -- project no.\ 511713970.

	\FloatBarrier
	\printbibliography

	\allowdisplaybreaks[1]
	\appendix
	
\end{document}